\documentclass{tlp}
\usepackage{graphicx}

\RequirePackage{amsmath, amssymb, latexsym}

\newtheorem{example}{Example}[section]
\newtheorem{definition}{Definition}[section]
\newtheorem{theorem}{Theorem}[section]

\newtheorem{procedure}{Procedure}[section]

\def\LDL{\mbox{$\cal LDL$}}
\def\bldl{\smallskip\[\tt\begin{array}{ll}}
\def\cldl{\[\tt\begin{array}{ll}}
\def\eldl{\end{array}\]\rm}
\def\prule#1#2{\tt #1 \leftarrow & \tt #2 \\}
\def\pfact#1#2{\tt #1 & \tt #2 \\}
\def\pbody#1#2{\tt #1 & \tt #2 \\}

\def\LDLXX{\mbox{$\cal LDL$}++}
\def\LDL{\mbox{$\cal LDL$}}
\def\inv{\vspace{0.0cm}}
\def\back{\hspace*{-2.0cm}}
\def\bk1{\hspace*{-1. cm}}

\def\<{\mbox{$\langle$}}
\def\>{\mbox{$\rangle$}}
\def\brack#1{\langle{#1}\rangle}
\def\etal{{\em et~al.}}\relax
\author[F. Arni and others]{FAIZ ARNI\\
InferData Corporation, 8200 N. MoPac Expressway, Austin, TX 78759, USA
\\ \email{farni@inferdata.com}
\and
KAYLIANG ONG\\ Trilogy Inc., 5001 Plaza on the Lake, Austin, TX 78746, USA\\ \email{kayliang.ong@trilogy.com}
\and
SHALOM TSUR\\ BEA Systems, \ 2315 N. First Street, San Jose, CA 95131, USA\\
\email{dicktsur@bea.com }
\and
HAIXUN WANG\\
IBM T. J. Watson Research Center, 30 Saw Mill River Rd.,
 Hawthorne, NY 10532, USA \\ \email{haixun@us.ibm.com}
\and CARLO ZANIOLO\\
Computer Science Department, University of California,  Los Angeles, CA, 90095, USA\\
\email{zaniolo@cs.ucla.edu}
}

\title{
The Deductive Database System \bf {\LDL}++ }

\jdate{January 2002}
\pubyear{2002}
\pagerange{\pageref{firstpage}--\pageref{lastpage}}

\begin{document}

\label{firstpage}

\maketitle

\begin{abstract}

This paper describes the {\LDL}++ system and  the research
advances that have enabled its design and development. We begin by
discussing the new nonmonotonic and nondeterministic constructs
that extend the functionality of the {\LDL}++ language, while
preserving its model-theoretic and fixpoint semantics. Then, we
describe the execution model and the open architecture designed to
support these new constructs and to facilitate the integration
with existing DBMSs and applications. Finally, we describe the
lessons learned by using {\LDL}++ on various tested applications,
such as middleware and datamining.
\end{abstract}


\section{Introduction}

The {\LDL}++ system, which was completed at UCLA in the summer of
2000, concludes a research project that was started at MCC in 1989
in response of the lessons learned from of its predecessor, the
{\LDL} system. The {\LDL} system, which was completed 1988,
featured many technical advances in language design  \cite{NaTs},
and implementation techniques \cite{Ceta}. However, its deployment
in actual applications \cite{Ts1,Ts2} revealed many problems and
needed improvements, which motivated the design of the new
{\LDL}++ system. Many of these  problems were addressed in the
early versions of the {\LDL}++ prototype that were built at MCC in
the period 1990--1993; but other problems, particularly limitations
due to the stratification requirement,  called for advances on
nonmonotonic semantics, for which solutions were discovered and
incorporated into the system over time---till the last version
(Version 5.1) completed at UCLA in the summer of 2000.

In this paper, we will concentrate on
the most  innovative and distinctive features of {\LDL}++, which can be
summarized as follows:

\begin{itemize}
\item
Its new language constructs designed to extend the expressive power of the
language, by allowing negation and aggregates in recursion, while retaining
the declarative semantics of Horn clauses,

\item
Its execution model designed to support (i) the new  language
constructs, (ii) data-intensive applications via tight coupling
with external databases, and (iii) an open architecture for
extensibility to new application domains,

\item
Its extensive application testbed designed to evaluate the
effectiveness of deductive database technology on
data intensive applications
and new domains, such as middleware and data mining.
\end{itemize}

\section{The Language}
A challenging research objective pursued by {\LDL}++ was that of
extending the expressive power of logic-based languages beyond
that of {\LDL} while retaining a fully declarative model-theoretic
and fixpoint semantics. As many other deductive database systems
designed in the 80s~\cite{Minker96}, the old {\LDL} system
required programs to be stratified with respect to nonmonotonic
constructs such as negation and set aggregates~\cite{RaUl95}.
While stratification represented a major step forward in taming
the difficult theoretical and practical problems posed by
nonmonotonicity in logic programs, it soon became clear that it
was too restrictive for many applications of practical importance.
Stratification makes it impossible to support efficiently even
basic applications, such as Bill of Materials and optimized
graph-traversals, whose procedural algorithms express simple and
useful generalizations of transitive closure computations. Thus,
deductive database researchers have striven to go beyond
stratification and allow negation and aggregates in the recursive
definitions of new predicates.
{\LDL}++ provides a comprehensive solution to
this complex problem by the fully integrated notions
of (i) $\tt choice$, (ii) User Defined Aggregates (UDAs), and (iii)
XY-stratification. Now, XY-stratification generalizes stratification
to support negation and (nonmonotonic) aggregates in recursion.
However, the choice construct (used to express functional dependency
constraints) defines mappings that, albeit nondeterministic, are
monotonic and can thus be used freely in recursion.
Moreover, this construct makes it possible
to provide a formal semantics to the notion of user-defined
aggregates (UDAs), and  to identify a special class of UDAs
that are monotonic \cite{ZWang99};
therefore, the {\LDL}++
compiler recognizes monotonic UDAs and allows their unrestricted
usage in recursion. In summary, {\LDL}++ provides a two-prong
solution to the nonmonotonicity problem, by (i) enlarging the class
of logic-based constructs that are monotonic (with constructs such
as choice and monotonic aggregates), and (ii) supporting
XY-stratification for hard-core nonmonotonic
constructs, such  as negation and nonmonotonic aggregates.

These new constructs of {\LDL}++ are fully integrated with all
other constructs, and easy to learn and use. Indeed, a user needs
not to know abstract semantic concepts, such as stable models or
well-founded models; instead, the user only needs to follow simple
syntactic rules---the same rules that are then checked by the
compiler. In fact, the semantic well-formedness of {\LDL}++
programs can be checked at compile time---a critical property of
stratified programs that was lost in later extensions, such as
modular stratification \cite{Ross94}. These new constructs are
described next.

\subsection{Functional Constraints}
\label{sec:choice}
Say that we have a database containing the
relations $\tt student(Name,  Major, Year)$
and $\tt professor(Name, Major)$.
In fact, let us take a toy example that only has
the following facts\footnote{We follow the standard convention
of using upper case initials to denote variables; lower case
initials and strings enclosed in quotes denote constants.}

\begin{eqnarray*}
\tt student('Jim Black', ee, senior). & \hspace*{1 cm} & \tt professor(ohm, ee). \\
                       &              & \tt professor(bell, ee).
\end{eqnarray*}

Now, the rule is that the major of a student must match his/her
advisor's major area of specialization.
Then, eligible advisors can be computed as
follows:

\bldl
\prule{elig\_adv(S, P) } { student(S,  Majr, Year),\ professor(P, Majr).}
\eldl

This yields

\bldl
\pfact{elig\_adv('Jim Black', ohm).} {}
\pfact{elig\_adv('Jim Black', bell).} {}
\eldl

But, since a  student can only have one advisor, the goal $\tt
choice((S),(P))$ must be added to our rule to force the selection  of a
unique advisor  for each student:

\begin{example} {\em Computation of unique advisors by a choice rule}
\label{advex} \bldl \prule{ actual\_adv(S, P) } {student(S, Majr,
Yr),\ professor(P, Majr),} \pbody{}      {choice((S),(P)).} \eldl
\end{example}

The goal $\tt choice((S),(P))$ can also be viewed as
enforcing a {\em functional dependency} (FD)  ${\tt S} \rightarrow {\tt P}$
on the results produced by the rule;
thus, in $\tt actual\_adv$, the second column (professor name) is
functionally dependent on the first one (student
name). Therefore, we will refer to {\tt S} and  {\tt P}, respectively,
as the left side and the right side of this FD,
and of the choice goal defining it.
The right side of a choice goal cannot be empty, but its
left side can be empty, denoting that all tuples produced
must share the same values for the right side attributes.

The result of  the rule of Example~\ref{advex}
is {\em nondeterministic}:
it can either return a singleton relation containing
the tuple $\tt ('Jim Black', ohm)$, or one
containing the tuple $\tt ('Jim Black', bell)$.

A program where the rules contain choice goals is called
a {\em choice program}.
The semantics of a choice program $P$ can be defined
by transforming $P$ into a program with negation,
$foe( P )$, called  the {\em first order equivalent} of $P$. Now,
$foe( P )$ exhibits a multiplicity of stable models,
each obeying the FDs defined
by the choice goals; each such stable model
corresponds to an alternative set of answers for  $P$ and
is called a {\em choice model} 
for $P$.
The first order equivalent of Example~\ref{advex} is as follows:

\begin{example} {The first order equivalent for Example~\ref{advex}}
\label{order}
\bldl
\prule{actual\_adv(S,P)} { student(S, Majr, Yr),\ professor(P, Majr),}
\pbody{}{chosen(S,P).}
\prule{chosen(S, P)    } { student(S, Majr, Yr),\ professor(P, Majr),}
\pbody{}{\neg diffChoice(S,P).}
\prule{diffChoice(S,P) } { chosen(S,P'),\ P \neq P'.}
\eldl
\noindent $\mbox{}$
\end{example}
This can be read as a statement that a professor will be assigned
to a student whenever a different professor has not been assigned to
the same student. In general, $foe( P )$  is defined as follows:

\begin{definition}\label{sv}
Let $P$ denote a program with choice rules: its
first order equivalent $foe(P)$ is
obtained by the following transformation. Consider
a choice rule $r$ in $P$:
\[
r : A \leftarrow B(Z),\ choice((X_1), (Y_1)),\ \ldots,\ choice((X_k), (Y_k)).
\]
where,

\begin{enumerate}
\item [(i)] $B(Z)$ denotes the conjunction of all the goals of $r$ that
are not choice goals, and

\item [(ii)] $X_i , \ Y_i , \ Z$, $1 \leq i \leq k$,
denote vectors of variables occurring in the body of $r$ such that
$X_i \cap Y_i = \emptyset$ and $X_i,Y_i \subseteq Z$.
\end{enumerate}
Then, $foe(P)$ is constructed from $P$ as follows:

\inv \begin{enumerate}
\item
Replace $r$ with a rule \ $r'$ obtained
by substituting the choice goals with the atom  $chosen_r(W)$:
\[
r' : A \leftarrow B(Z), \ chosen_r(W) .
\]
where $W \subseteq Z$ is the list of all variables appearing in choice
goals, i.e., $W = \bigcup_{1 \leq j \leq k} X_j \cup Y_j$.\vspace{0.2cm}
\item Add the new rule
\[ chosen_r (W) \leftarrow B(Z), \ \neg diffChoice_r (W). \]
\item
For each choice atom $choice((X_i),(Y_i))$ ($1 \leq i \leq k$),
add the new rule
\[
diffChoice_r (W) \leftarrow chosen_r(W'),  \ Y_i \neq Y'_i.
\]
where (i) the list of variables $W'$ is derived from $W$ by
replacing each $A \not\in X_i$ with a new variable $A'$ (i.e., by
priming those variables), and (ii)
$Y_i \neq Y '_i$
denotes the
inequality of the vectors;  i.e.,
$Y_i \neq Y '_i$ is
true when
for some variable  $A \in Y_i$ and its primed
counterpart $A' \in Y' _i$, $A \neq A '$.
~\hfill~ $\mbox{}$
\end{enumerate}
\end{definition}

\subsubsection*{Monotonic  Nondeterminism}
\inv
\begin{theorem} Let $P$ be a positive program with
choice rules. Then
the following properties hold~\cite{GPSZ}:

\inv \begin{itemize}
\item
$foe(P)$ has one or more total stable models.
\inv \item
The {\em chosen} atoms in each stable model of $foe(P)$
obey the FDs defined by the choice goals.
\end{itemize}
\end{theorem}

Observe that the $foe(P)$ of a program with choice does
not have total well-founded models; in fact, for our
Example~\ref{advex},  the well-founded model
yields undefined values for advisors. Therefore,
the choice construct can express nondeterministic
semantics, which can be also expressed by
stable models, but not by well-founded models. On the other
hand, the choice model avoids the exponential complexity which
is normally encountered under stable model semantics.
Indeed, the computation of stable models is $\cal NP$-hard
\cite{Sch92}, but the computation of choice models for positive
programs can be performed in polynomial time with respect to the
size of the database. This, basically, is due to the monotonic
nature of the choice construct that yields a simple fixpoint
computation for programs with choice \cite{JCSS}. Indeed, the use
of choice rules in positive programs preserves their monotonic
properties. A program $P$ can be viewed as  consisting of two
separate components: an extensional component (i.e., the database
facts), denoted $edb(P)$, and an intensional one (i.e., the
rules), denoted $idb(P)$. Then, a positive choice program defines
a monotonic multi-valued mapping from $edb(P)$ to $idb(P)$, as per
the following theorem proven in \cite{JCSS}:

\begin{theorem} \label{th:jcss}
  Let $P$ and $P'$ be two positive choice programs where
  $idb(P')=idb(P)$ and $edb(P') \supseteq edb(P)$. Then, if
  $M$ is a choice model for $P$, then, there exists a choice
  model $M'$ for $P'$ such that $M' \supseteq M$.
\end{theorem}

Two  concrete semantics are possible for choice programs:
one is an all-answers semantics,  and the
other is the semantics under which any answer will do---don't care nondeterminism.
While an all-answers semantics for choice is not without interesting
applications \cite{ILPS97}, the single-answer semantics
was adopted by {\LDL}++, because this is effective at
supporting  {\em DB-PTime} problems \cite{AHV95}. Then, we see that
Theorem 2 allows us to compute results incrementally as it is done
in differential fixpoint computations; in fact, to
find an answer, a program with choice can be implemented
as an ordinary program, where the choice predicates are memorized
in a table; then newly derived atoms that violate the choice
FDs are  simply discarded, much in the same way as duplicate
atoms are discarded during a fixpoint computation.
Thus positive choice programs represent a class of logic
programs that are very well-behaved from both a  semantic
and a computational viewpoint. The same can be said for
choice programs with stratified negation that are defined next.

\begin{definition} Let $P$ be a  choice program with negated goals.
Then, $P$ is said to be stratified
when the program obtained from $P$
by removing its choice goals is stratified.
\end{definition}

The stable models for a  stratified choice program $P$
can be  computed using an
{\em iterated choice fixpoint} 
procedure that directly extends the iterated
fixpoint procedure for programs with
stratified negation~\cite{Prz88,ZaCe97}; this is summarized next.
Let $P_i$, denote the rules of $P$ (whose head is) in stratum
$i$, and let $ {P_i}^*$ be the union of $ P_j, \ j \leq i$. Now, if
$M_i$ is a stable model for ${P_i}^*$, then every
stable model for $M_i \cup P_{i+1}$ is a stable model
for the program ${P^*_{i+1}}$.
Therefore, the stable models of
stratified choice programs can be computed by modifying
the iterated fixpoint
procedure used for stratified programs
so that choice models (rather than the least models)
are computed for strata containing
choice rules~\cite{GiaMa98}.


\subsection*{The Power of Choice} \label{choicealgos}
The expressive power of choice was studied in \cite{JCSS}, where it
was shown that stratified Datalog with choice can express all
computations that are polynomial in the size of the database (i.e.,
DB-PTIME queries~\cite{AHV95}). Without choice, DB-PTIME
cannot be expressed in stratified Datalog,
unless a predefined total
order is assumed for the universe, an assumption that
would violate the {\em genericity} principle \cite{AHV95}.
In terms of  computational power,
non-determinism and order fulfill a similar function~\cite{AHV95};
in fact, the application of choice can also
be viewed as non-deterministically and incrementally
generating a possible order on the universe---an order that
is made explicit by the
predicate $\tt chain$ discussed in Example~\ref{ex4}.

Before moving to Example~\ref{ex4}, however,
we would like to observe that
the version of choice supported
in {\LDL}++ is more  powerful than other nondeterministic
constructs, such as the witness operator~\cite{AHV95}, and an earlier
version of choice proposed in~\cite{KriNaq88} (called static choice
in \cite{JCSS}).
For instance, the following query cannot be expressed in standard
Datalog (since the query is nondeterministic) nor it can be
expressed by the early version of  choice~\cite{KriNaq88} or by
the witness construct ~\cite{AHV95}. These early constructs
express nondeterminism in nonrecursive programs, but suffer from
inadequate expressive power in recursive programs~\cite{JCSS}. In
particular, they cannot express the query in Example \ref{span1}.

\begin{example}\label{span1}
{\em Rooted spanning tree.}
We are given an
undirected graph where an
edge joining two nodes, say $x$ and $y$, is represented by
the pair $g(x,y)$ and $g(y,x)$. Then,
a spanning tree in this graph, starting from the source node $\tt a$,
can be constructed by the following program:

\bldl
\pfact{ st(root, a). } { }
\prule{ st(X, Y)        } { st(\_,X), \ g(X, Y), \ Y \neq a, \ Y \neq X, }
\pbody{}{choice((Y),(X)). }
\eldl
\noindent
To illustrate the presence of
multiple total choice models for this program,
take a simple graph consisting of the following arcs:

\bldl
\pfact{ g(a, b). } { g(b, a).}
\pfact{ g(b, c). } { g(c, b).}
\pfact{ g(a, c). } { g(c, a).}
\eldl
\end{example}

After the exit rule adds $\tt st(root, a)$,
the recursive rule could add  $\tt st(a, b)$ and
$\tt  st(a, c)$, along with the two tuples
$\tt  chosen(a,b)$ and $\tt chosen(a,c)$
in the $\tt chosen$ table.
No further arc can be added after those,
since the addition of $\tt st(b, c)$ or
$\tt st(c, b)$ would violate
the FD that follows from $\tt choice((Y),(X))$
enforced through the $\tt chosen$ table.
However, since
$\tt st(root,$ $\tt a)$ was produced by the first rule (the exit rule),
rather than the second rule (the recursive choice rule), the
table $\tt chosen$ contains no tuple with second argument equal to the
source node $\tt a$.
Therefore, to avoid the addition of
$\tt st(c, a)$ or $\tt st(b, a)$, the goal
$\tt Y \neq a$ was added to the recursive rule.

By examining all possible solutions, we
conclude that this program has three different choice models,
for which we list only the $\tt st$-atoms, below:

\bldl
1.\ \pfact{ \ st(a, b), } { st(b, c).}
2.\ \pfact{ \ st(a, b), } { st(a, c).}
3.\ \pfact{ \ st(a, c), } { st(c, b).}
\eldl
\noindent $\mbox{}$

In addition to supporting {\em nondeterministic} queries, the
introduction of the choice extends the power of Datalog for {\em
deterministic} queries.
This can be illustrated by the following
choice program that places the elements of a relation $\tt d(Y)$ into
a chain, thus establishing a random total order on these elements; then
checks if the last element in the chain is even.

\begin{example}\label{ex4}
{The odd parity query by arranging the elements of a set in a chain.
The elements of the set are stored by means of facts of the form $\tt
d(Y)$.}
 \bldl \pfact{ chain(nil, nil). } { }
 \prule{ chain(X,Y) } {chain(\_,X), \ d(Y),}
 \pbody{} {choice((X),(Y)), \ choice((Y),(X)). \ \ \ \ }
\prule{odd(X)} {chain(nil,X).}
\prule{odd(Z)} {odd(X), chain(X,
Y),chain (Y,Z).}
\prule{isodd} {odd(X), \neg chain(X, Y).} \eldl
\noindent $\mbox{}$
\end{example}

Here  $\tt chain(nil, nil)$ is the root of a chain linking all the
elements of $\tt d(Y)$---thus inducing a total order on elements
of $\tt d$.

The negated goal in the last rule defines the last element in the chain.
Observe that the final $\tt isodd$ answer does not depend on
the particular chain constructed; it only depends on its length
that is equal to the cardinality of the set. Thus stratified Datalog
with choice can express deterministic queries,
such as the parity query, that cannot be expressed in
stratified Datalog without choice \cite{AHV95}.


The parity query cannot be expressed in  Datalog with stratified negation
unless we assume that the underlying universe
is totally ordered---an assumption that
violates the data independence principle of
{\em genericity} \cite{AHV95}.
The benefits of this added expressive power in real-life applications
follows from the fact that the chain program
used in Example \ref{ex4}, above, to compute the odd parity query
can be used to stream through the elements
of a set one by one,
and compute arbitrary aggregates on them.
For instance, to  count the cardinality of the set $\tt d(Y)$ we can write:
\inv\bldl
\prule{mcount(X,1)} {chain(nil,X).}
\prule{mcount(Y, J1)} {mcount(X, J), chain(X, Y), J1=J+1.}
\prule{count(J)} {mcount(X,J), \neg chain(X, Y).}
\inv \eldl
The negated goal in the last rule qualifies the
element(s) $\tt X$ without a successor in the chain,
i.e., $\tt X$ for which
$\tt \neg chain(X,Y)$ holds for all $\tt Y$s.
Therefore,   {\tt count}
is defined by a program containing
(and stratified with respect to)
negation; thus, if $\tt count$ is then used as  a builtin aggregate,
the stratification requirement must be enforced upon every
program that uses  $\tt count$.

However, if we seek to determine if the base relation
$\tt d(Y)$ has more than $14$ elements, then
we can use the $\tt mcount$ aggregate instead of $\tt count$,
as follows:
\inv
\bldl
\prule{morethan14} {mcount(\_, J), J>14. }\inv
\eldl

Now,
$\tt mcount$ is what is commonly known as an {\em online}
aggregate \cite{hellerstein}:
i.e., an aggregate that produces {\em early returns} rather than final
returns as traditional aggregates. The use of $\tt mcount$ over $\tt count$
offers clear performance benefits; in fact, the computation
of  $\tt morethan14$ can be terminated after 14 items, whereas
the application of
$\tt count$ requires visiting all the items in the chain.
From a logical viewpoint, the benefits are even greater,
since $\tt count$ is no longer needed and the rule defining it
can be eliminated---leaving us
with the program defining $\tt mcount$, which is free of
negation. Thus,
no restriction is needed when  using $\tt mcount$ in
recursive programs; and indeed, $\tt mcount$ (and $\tt morethan14$)
define monotonic mappings in the lattice of set-containment.

In summary, the use of choice
led us to (i) a simple and general definition of the concept of
aggregates, including user defined aggregates (UDAs), and (ii)
the identification of a special subclass of UDAs
that are free from the yoke of stratification, because
they are monotonic. This topic is further discussed
in the next section.

\subsection{User Defined Aggregates}\label{sec:udas}
\inv
The importance of aggregates in deductive databases has been
recognized for a long time~\cite{RosSa97,VG93,Kemp98}. In
particular, there have been several attempts to overcome the
limitations placed on the use of aggregates in programs because of
their nonmonotonic nature~\cite{Filk96}. Of particular interest is
the work presented in \cite{RosSa97}, where it shown that rules
with aggregates often define monotonic mappings in special
lattices---i.e., in lattices different from the standard
set-containment lattice used for $T_P$. Furthermore, programs with
such monotonic aggregates can express many interesting
applications \cite{RosSa97}. Unfortunately, the lattice that makes
the aggregate rules of a given program monotonic is very difficult
to identify automatically ~\cite{VG93}; this problem prevents the
deployment of such a notion of monotonicity in real deductive
database systems.

A new wave of decision support applications has recently underscored
the importance of aggregates and the need for a wide range of
new aggregates~\cite{han}.
Examples include rollups and datacubes for OLAP applications,
running aggregates and window aggregates in time series analysis,
and special versions of standard aggregates used to
construct classifiers or association rules
in datamining \cite{han}. Furthermore,
a new form of aggregation,
called online aggregation, finds many uses in
data-intensive applications \cite{hellerstein}.
To better serve this wide new assortment
of applications
requiring specialized aggregates, a deductive database system
should support User Defined Aggregates (UDAs).
Therefore, the new  {\LDL}++
system  supports powerful
UDAs, including online aggregates and monotonic
aggregates, in a simple rule-based
framework built on formal logic-based
semantics.

In {\LDL}++ users can define a new aggregate by writing
the {\tt single}, {\tt multi}, and {\tt
freturn} rules (however, {\tt ereturn} rules can be
used to supplement or
replace {\tt freturn} rules).
The {\tt single} rule defines the computation for the
first element of the set (for instance $\tt mcount$ has its second argument
set to $1$), while {\tt multi} defines the induction step whereby the
value of the aggregate
on a set of $n+1$ elements is derived from the aggregate value of
the previous set with $n$ elements and the value of
$(n+1)^{th}$ element itself. A
unique aggregate name is used as the first argument in
the head of these rules to eliminate any interference between the
rules defining different aggregates. For instance, for computing
averages we must compute both the count and the
sum of the elements seen so far:
\bldl
\pfact{single(avg, Y,  cs(1, Y)).} {}
\prule{multi(avg, Y, cs(Cnt, Sum), cs(Cnt1, Sum1))}{}
\pfact {}{\hspace*{-2.8cm}Cnt1=Cnt+1, Sum1=Sum+Y.}
\eldl

Then, we write a {\tt freturn}  rule that upon
visiting the final element in $\tt d(Y)$
produces the ratio of sum over count, as follows:

\bldl
\prule {freturn(avg, Y, cs(Cnt, Sum), Val)} {Val= Sum/Cnt.}
\eldl

After an aggregate is defined by its
$\tt single$, $\tt multi$,
$\tt ereturn$ and/or  $\tt freturn$ rules,
it can be invoked and used in
other rules. For instance, our
the newly defined $\tt avg$ can be invoked as follows:

\bldl
\prule{p(avg \< Y \>)} {d(Y).}
\eldl
Thus {\LDL}++ uses the special notation of
pointed brackets, in the head of rules,  to denote the application of
an aggregate. This syntax, that has been adopted by
other languages \cite{Rama93},
also supports an implicit `group by' construct, whereby
the aggregate arguments in the head are implicitly grouped by
the other arguments in the head. Thus,
to find the average salary of employees grouped by department a
user can write the following rule:
\bldl
\prule{davg(DeptNo, avg \< Sal \>)} { employee(Eno, Sal, DeptNo).}
\eldl

The formal semantics of UDAs
was introduced in \cite{ZWang99} and is described in
the Appendix: basically, the aggregate invocation rules
and the aggregate definition rules are rewritten into an equivalent
program that calls on the chain predicate defined as in Example \ref{ex4}.
(Naturally, for the sake of efficiency, the {\LDL}++
system shortcuts the full rewriting used to define
their formal semantics, and implement the
UDAs by a more direct implementation.)

{\LDL}++  UDAs have also been extended to support online
aggregation~\cite{hellerstein}. This is achieved by using  $\tt
ereturn$ rules in the definition of UDAs, to  either supplement,
or replace $\tt freturn$ rules.


For example, the computation of averages normally produces an
approximate value long before the whole data set is visited. Then,
we might want to see the average value obtained so far every 100
elements. Then, the following rule will be added:

\bldl
\pfact {ereturn(avg, X,(Sum,Count), Avg) \leftarrow
\hspace*{-2cm}} {}
\pbody{}{ Count \ mod \ 100 = 0, Avg =
Sum/Count.} \eldl

Thus the $\tt ereturn$ rules produce {\em early returns}, while
the {\tt freturn} rules produce final returns.

As second example, let us consider the well-known problem
of coalescing after temporal projection
in temporal databases \cite{ZaCe97}. For instance in Example 5, below,
after projecting out from the employee relation the salary column,
we might have a situation where the same
$\tt Eno$ appears in tuples where their valid-time
intervals overlap; then these intervals must be coalesced. Here,
we use closed intervals represented by the pair $\tt (From, To)$
where $\tt From$ is the  start-time, and $\tt To$ is the end-time.
Under the assumption that tuples are sorted by increasing
start-time, then we can use a special $\tt coales$ aggregate
to perform the task in one pass through the data.

\begin{example}{Coalescing overlapping intervals sorted by
start time.} \label{coalesc}
\bldl \prule{empProj(Eno,
coales\<(From, To)\>)}{emp(Eno,\_,\_,(From, To)).}
\eldl \inv
 \cldl
\pfact{single(coales, (Frm, To), (Frm, To)).}{}
\prule{multi(coales, (Nfr, Nto), (Cfr, Cto),(Cfr, Nto))} {}
\pbody{} {\hspace*{-2.4cm} Nfr <= Cto, Nto > Cto.}
\prule{multi(coales, (Nfr, Nto), (Cfr, Cto), (Cfr, Cto))}{}
\pbody{} {\hspace*{-2.4cm} Nfr <= Cto, Nto <= Cto.}
\prule{multi(coales, (Nfr, Nto), (Cfr, Cto), (Nfr, Nto))} { Cto < Nfr. }
\eldl

\cldl
\prule{ereturn(coales, (Nfr, Nto), (Cfr, Cto), (Cfr, Cto))} { Cto < Nfr. }
\pfact{freturn(coales, \_,  LastInt, LastInt).}{} \eldl
\end{example}

Since the input intervals are ordered by
their start time,
the new interval $\tt (Nfr, Nto)$ overlaps
the current interval $\tt (Cfr, Cto)$
when $\tt Nfr \leq Cto$; in this situation, the
two intervals are merged into one that begins at $\tt Cfr$ and
ends with the larger of $\tt Nto$ and $\tt Cto$.
When, the new interval does not overlap with the current
interval, this is returned by the ereturn rule,
while the new interval becomes the current
one (see the last multi rule).

Let $P$ be a program. A rule $r$ of $P$ whose head contains aggregates  is
called an {\em aggregate rule}. Then, $P$ is said to be
stratified w.r.t. aggregates when for each aggregate  rule $r$ in
$P$, the stratum of $r$'s head predicate is strictly higher than the
stratum of each predicate in the head of $r$. Therefore,
the previous program is stratified with respect to $\tt coales$ which is
nonmonotonic since it uses both early returns and final returns.

While, programs stratified with respect to aggregates can be used
in many applications, more advanced applications
require the use of aggregates in more general settings. Thus,
{\LDL}++ supports
the usage of arbitrary aggregates in XY-stratified programs, which
will be discussed in Section 3. Furthermore {\LDL}++ supports the
monotonic aggregates that can be used freely in recursion.

\subsection*{Monotone Aggregation}
An important result that follows from the formalization
of the semantics of UDAs \cite{ZWang99} (see also Appendix),
is that UDA defined without  final
return rules, i.e., no {\tt freturn} rule, define monotonic
mappings, and can thus be used without restrictions in
the definition of recursive predicates.
For instance, we will next define a continuous count that
returns the current count after
each new element (thus final returns are here omitted
since they are redundant).
\inv
\bldl
\pfact{single(mcount, Y, 1).}{}
\prule{multi(mcount, Y, Old, New)} { New=Old+1.}
\prule{ereturn(mcount, Y, Old, New)} { New=Old+1.}
\eldl

Monotonic aggregates allow
us to express the following two examples taken
from \cite{RosSa97}.

\paragraph{Join the Party}
Some people will come to the party no matter what, and their
names are stored in a $\tt sure(Person)$ relation.  But
others will join only after they know that at least $K=3$ of
their friends will be there.  Here, $\tt friend(P, F)$
denotes that $\tt F$ is a friend of person $\tt P$.

\bldl
\prule{willcome(P)}{sure(P).}
\prule{willcome(P)}{c\_friends(P, K), K \geq 3.}
\prule{c\_friends(P, mcount\brack{ F}) }{willcome(F), friend(P, F).}
\eldl

Consider now a computation of these rules on
the following database.

\bldl
\pfact{friend(jerry, mark). }{sure(mark).}
\pfact{friend(penny, mark). }{sure(tom).}
\pfact{friend(jerry, jane). }{sure(jane).}
\pfact{friend(penny, jane). }{}
\pfact{friend(jerry, penny). \hspace{1.4cm} }{}
\pfact{friend(penny, tom). }{}
\eldl

Then, the basic semi-naive computation yields:
$$ \tt willcome(mark), willcome(tom), willcome(jane),$$
$$ \tt c\_friends(jerry, 1), c\_friends(penny, 1), c\_friends(jerry, 2),$$
$$\tt c\_friends(penny, 2), c\_friends(penny, 3), willcome(penny),$$
$$\tt c\_friends(jerry, 3), willcome(jerry).$$

This example illustrates how the standard semi-naive computation can be
applied to queries containing monotone UDAs.
Another interesting example is
transitive ownership and control of corporations.
\paragraph{Company Control}
Say that $\tt owns(C1, C2, Per)$ denotes
the percentage of shares that corporation
$\tt C1$ owns of corporation $\tt C2$.
Then, $\tt C1$ controls $\tt C2$ if it owns
more than, say, $50 \%$ of its shares. In general, to decide
whether $\tt C1$ controls $\tt C3$ we must also add the shares owned
by corporations, such as $\tt C2$, that are controlled by $\tt C1$.
This yields the transitive control rules defined
with the help of a continuous sum  aggregate that returns the
partial sum for each new element:

\inv\bldl
\prule{control(C, C)}{owns(C, \_, \_).}
\prule{control(Onr, C) }{towns(Onr, C, Per), Per>50.}
\prule{towns(Onr, C2, msum\brack{ Per}) }{control(Onr, C1), owns(C1, C2, Per).}
\eldl
\cldl
\pfact{single(msum, Y, Y).}{}
\prule{multi(msum, Y, Old, New)} { New=Old+Y.}
\prule{ereturn(msum, Y, Old, New)} {New=Old+Y.}
\eldl

Thus, every company controls itself, and a company
$\tt C1$ that has transitive
ownership of more than $50 \%$ of  $\tt C2$'s shares controls $\tt C2$.
In the last rule, $\tt towns$ computes transitive ownership
with the help of $\tt msum$ that adds up the shares of controlling
companies. Observe that any pair $\tt (Onr,C2)$ is added at most
once to $\tt control$, thus the contribution of $\tt C1$ to $\tt Onr$'s
transitive ownership of $\tt C2$  is only accounted once.

\paragraph{ Bill-of-Materials 
(BoM) Applications}
BoM applications represent an important application
area that requires aggregates in recursive rules.
For instance, let us say  that $\tt assembly(P1, P2, QT )$
denotes that $\tt P1$ contains part $\tt P2$ in quantity $\tt QT$.
We also have
elementary parts described by the relation
$\tt basic\_part(Part, Price)$. Then, the following program computes
the cost of a part as the sum of the cost of the basic parts
it contains:

\bldl
\prule{part\_cost(Part, O, Cst)} {\back basic\_part(Part, Cst).}
\prule{part\_cost(Part,  mcount\brack{ Sb}, msum\brack{ MCst})}{}
\pbody{}{\back \back part\_cost(Sb, ChC, Cst), prolfc(Sb, ChC),}
\pbody{} {\back \back assembly(Part, Sb, Mult), MCst=Cst*Mult.}
\eldl

Thus, the key condition in the body of the second rule is that a
subpart $\tt Sb$ is counted in $\tt part\_cost$ only when all $\tt
Sb$'s children have been counted.  This occurs when the number of $\tt
Sb$'s children counted so far by $\tt mcount$ is equal to the out-degree
of this node in the graph representing $\tt assembly$.
This number is kept in
the prolificacy table, $\tt prolfc(Part, ChC)$, which can be computed
as follows:

\bldl
\prule{prolfc(P1, count\brack{ P2})}{assembly(P1, P2, \_).}
\prule{prolfc(P1, 0)}{basic\_part(P1, \_).}
\eldl

Therefore the simple and general solution of the monotonic
aggregation problem introduced by {\LDL}++ allows the concise
expression of many interesting algorithms.
This concept
can also be  extended easily
to SQL recursive queries,
as discussed in \cite{icde2k} where additional
applications are also discussed.

\subsection{Beyond Stratification}

The need to go beyond stratification has motivated much
recent research. Several deductive database systems have addressed
it by supporting the notion of modular
stratification \cite{Ross94}. Unfortunately, this approach suffers
from poor usability, since the existence of a  modular stratification
for a program can depend on its extensional information (i.e., its
fact base) and, in general, cannot be checked without executing the
program. The standard notion of
stratification is instead much easier to use,
since it provides a simple criterion for the
programmer to follow and for the compiler to use when validating the
program and optimizing its execution.
Therefore, {\LDL}++ has introduced
the notion of XY-stratified programs 
that preserves
the compilability and usability benefits of stratified programs
while achieving the expressive power of well-founded
models \cite{Kemp95}.  XY-stratified programs
are locally stratified explicitly by a temporal argument: thus,
they can be viewed as Datalog$_{1S}$ 
programs, which are known to provide a powerful
tool for temporal reasoning \cite{bacho94,ZaCe97},
or as Statelog 
programs that were used to model active
databases~\cite{LaLu98}.  The deductive  database system
Aditi \cite{Kemp98} also supports the closely related concept of
explicitly locally stratified programs,
which were shown to
be as powerful as well-founded
models, since they can express their
alternating fixpoint computation~\cite{Kemp95}.

For instance, the ancestors of {\tt marc}, with the number of generations
that separate them from {\tt marc}, can be computed using the
following program which  models the differential fixpoint
computation:

\begin{example} {Computing ancestors of Marc and their remoteness from Marc
using differential fixpoint approach.}
\label{deltagap}
\bldl
r_1: \pfact{delta\_anc(0, marc).}{}
r_2: \prule{delta\_anc(J+1 , Y)} {delta\_anc(J, X), parent(Y,X),}
\pbody{}{\neg all\_anc(J, Y).}
r_3:\prule{all\_anc(J+1 , X)}{ all\_anc(J, X).}
r_4:\prule{all\_anc(J, X) }{ delta\_anc(J,  X).}
\eldl
\end{example}

This program is locally stratified  by
the first arguments in $\tt delta\_anc$ and $\tt all\_anc$ that serve as
temporal arguments (thus +1 is a postfix successor function symbol,
much the same as $s(J)$ that denotes the successor of $J$ in
Datalog$_{1S}$ \cite{ZaCe97}).
The zero  stratum consists of atoms of nonrecursive
predicates such as $\tt parent$ and of atoms that unify with
$\tt all\_anc(0, X)$ or $\tt delta\_anc(0, X)$.
The $k^{th}$ stratum
consists of atoms of the form
$\tt all\_anc(k, X)$, $\tt delta\_anc(k, X)$.
Thus, the previous program is locally
stratified \cite{Prz88}, since the heads
of recursive rules belong to strata that are one
above those of their goals.
Alternatively, we can view the previous program as a
compact representation for
the stratified program obtained
by instantiating the temporal argument to
integers and attaching them to the predicate names, thus generating
an infinite sequence of unique names.

Also observe that the temporal arguments in rules
are either the same as, or one less than, the temporal argument in the head.
Then, there are two kinds of rules in our example:
(i) X-rules (i.e., a horizontal rules) where the
temporal argument in each  of their goals is the same
as that in their heads,
and (ii) Y-rules (i.e., a vertical rules) where the
temporal arguments in some of their goals are one less
than those in their heads.
Formally, let $P$ be a set of rules
defining mutually recursive predicates,
where each recursive predicate has a distinguished
temporal argument and every rule in $P$ is
either an X-rule or a Y-rule. Then, $P$ will be said to
be an XY-program. For instance, the program
in Example \ref{deltagap} is an XY-program,
where $r_4$ and $r_1$ are X-rules, while $r_2$ and $r_3$
are Y-rules.

A simple test can now be used
to decide whether an XY-program $P$ is locally
stratified. The test begins by labelling
all the head predicates in $P$ with the prefix `new'.
Then, the body predicates with
the same temporal argument as the head
are also labelled with the prefix `new',
while the others are labelled with
the prefix `old'.  Finally, the temporal arguments are dropped from
the program. The resulting program is called the
{\em bistate version} of $P$ and is denoted  $P_{bis}$.

\begin{example}{The bistate version of the program in Example \ref{deltagap}}
\label{bistate}
\bldl
\pfact{new\_delta\_anc( marc).}{}
\prule{new\_delta\_anc(Y)}
     {old\_delta\_anc(X), parent(Y,X),}
\pbody{}{\neg old\_all\_anc(Y).}
\prule{new\_all\_anc(X) }{ new\_delta\_anc(X).}
\prule{new\_all\_anc(X)}{ old\_all\_anc(X).}
\eldl
\end{example}

Now we have that~\cite{ZAO93}: 

\begin{definition}
Let $P$ be an XY-program. $P$ is said to be XY-stratified when
$P_{bis}$ is a stratified program.
\end{definition}

\begin{theorem}
Let $P$ be an XY-stratified program. Then $P$ is locally stratified.
\end{theorem}

The program of Example \ref{bistate} is stratified with the
following strata: $ \tt S_0 = \{ parent,$ $ \tt old\_all\_anc,$
$\tt old\_delta\_anc \}$,
$\tt S_1 = \{ new\_delta\_anc \}$, and
$ \tt S_2 = \{ new\_all\_anc \}$. Thus, the program
in Example \ref{deltagap} is locally stratified.

For an XY-stratified program $P$, the general
iterated fixpoint procedure \cite{Prz88}  used to compute
the stable model of locally stratified programs
\cite{ZAO93} becomes quite
simple; basically it reduces to a repeated computation over
the stratified program $P_{bis}$. For instance,
for Example \ref{bistate}
we compute
$\tt new\_delta\_anc$ from $\tt old\_delta\_anc$ and then
$\tt  new\_all\_anc $ from this. Then, the `old' relations are
re-initialized with the content of the 'new' ones so derived, and
the process is repeated.
Furthermore, since the temporal
arguments have been removed from this program, we need
to
\begin{enumerate}
\item store the temporal argument as an external fact
$\tt counter(T)$,
\item add a new goal ${\tt counter(I}_r)$
to each exit rule $r$ in  $P_{bis}$,
where ${\tt I}_r$ is the variable from the temporal
arguments of the original rule $r$, and
\item For each recursive predicate $\tt q$ add the rule:
\inv \bldl
\prule{q(J, X)}{ new\_q(X), counter(J).}\\
\eldl
\end{enumerate}
The program so constructed
will be called the {\em synchronized} bistate version of $P$, denoted
$syncbi(P)$.
For instance, to obtain the synchronized version of the program
in Example \ref{bistate}, we need to change the first rule to
\bldl
\prule{new\_delta\_anc( marc)}{counter(0).}
\eldl
since the temporal argument in the original exit rule was the
constant $0$. Then, we must add the following rules:

\inv \bldl
\prule{delta\_anc(J, X)}{ new\_delta\_anc(X), counter(J).}
\prule{all\_anc(J, X)}{ new\_all\_anc(X), counter(J).}\\
\eldl

Then, the iterated fixpoint computation for an XY-stratified
program can be implemented by the following procedure:

\inv
\begin{procedure} {\em Computing a stable model of
an XY-stratified program $P$:}
\label{xycomp}
\noindent
Add the fact $\tt counter(0)$. Then,
forever repeat the following two steps:

\inv
\begin{enumerate}
\item
Compute the stable model of $syncbi(P)$.

\item
For each recursive predicate $\tt q$,
replace  $\tt old\_q$ with $\tt new\_q$, computed in the previous
step. Then, increase the value of  $\tt counter$ by one.
\end{enumerate}
\end{procedure}

Since $syncbi(P)$ is stratified, we can then
use the iterated fixpoint computation to compute its
stable model.

Since each XY-stratified program is locally stratified \cite{Prz88},
it is guaranteed to have a unique stable model, which is also known
as its perfect model \cite{Prz88}. But the special syntactic structure
of XY-stratified programs allows an efficient computation of their
perfect models using Procedure 4;  moreover, in the actual
{\LDL}++ implementation, this computation is further improved
with the optimization techniques discussed next.
For instance,
the replacement of $\tt old\_q$ with $\tt new\_q$
described in the last step of Procedure \ref{xycomp}
becomes an operation 
of (small) constant cost when it is  implemented
by switching the pointers to the
relations. A second improvement concerns
{\em copy rules}, such as the last rule in Example \ref{deltagap}.
For instance
$r_3$ in Example 6 is a copy rule that copies the
new values of $\tt all\_anc$ from its old values.
Observe that
the body and the head of this rule are identical, except for
the prefixes $\tt new$ or $\tt old$,
in its bistate version (Example \ref{bistate}).
Thus, in order to
compute $\tt new\_ all\_anc$, we first execute
the copy rule by simply setting the pointer to $\tt new\_ all\_anc$
to point to $\tt old\_ all\_anc$---a constant-time operation.
Rule $r_4$ that adds tuples to $\tt new\_ all\_anc$
is then executed after $r_3$.

In writing XY-stratified programs, the user must also
be concerned with termination conditions, since e.g.,
a rule such as $r_3$ in Example \ref{deltagap} could, if left unchecked,
keep producing $\tt all\_anc$ results under
a new temporal argument, after $\tt delta$ becomes empty.
One solution to this problem is for the user
to add the goal $\tt delta\_anc(J, \_)$ to rule $r_3$.
Then, the computation $\tt all\_anc$ stops
as soon as no new  $\tt delta\_anc(J, \_)$ is generated.
Alternatively,
our program could be called from a  goal such as
$\tt delta\_anc(J, Y)$. In this case, if $r_2$
fails to produce any result for a value $\tt J$, no
more results can be produced at successive steps,
since $\tt delta\_anc(J, Y)$ is a positive goal of $r_2$.
The {\LDL}++ system is capable of
recognizing these situations, and it will
terminate the computation of Procedure \ref{xycomp} when
either condition occurs.

Example \ref{overlap} solves the coalescing problem
without relying on
tuples being sorted on their start-time---an assumption
made in Example \ref{coalesc}. Therefore, we
use the XY-stratified program
of Example~\ref{overlap}, which iterates over
two basic computation steps. The first step is defined by the
{\tt overlap} rule, which identifies
pairs of distinct intervals that overlap, where
the first interval contains the start of the second interval.
The second step consists of deriving a new interval
that begins at the start of the first interval, and ends at the later
of the two endpoints. Finally,
a rule $\tt final\_e\_hist$  returns the
intervals that do not overlap other intervals (after
eliminating the temporal argument).\\

\begin{example}{Coalescing  overlapping periods into
maximal periods after a projection}
\label{overlap}
\bldl
\prule{e\_hist(0,Eno, Frm, To) }{ emp\_dep\_sal(0,Eno, \_, \_, Frm, To).  }
\pfact{\lefteqn{\tt overlap(J+1, Eno, Frm1, To1, Frm2, To2)  \leftarrow} }{}
\pfact{}{e\_hist(J, Eno, Frm1, To1),}
\pfact{}{e\_hist(J, Eno, Frm2, To2),}
\pfact{}{Frm1 \leq Frm2, Frm2 \leq To1,}
\pfact{}{distinct(Frm1, To1, Frm2, To2).
}
\prule{e\_hist(J,Eno, Frm1, To)}{\! overlap(J,Eno, Frm1, To1, Frm2, To2),}
\pbody{}      {\! select\_larger(To1, To2, To).}
\eldl
\inv
\bldl
\prule{final\_e\_hist(J+1,Eno, Frm, To) }{ e\_hist(J,Eno, Frm, To),}
\pbody{}{\neg overlap(J+1,Eno, Frm, To, \_, \_).}
\eldl
\cldl
\prule{distinct(Frm1, To1,Frm2, To2)}{To1 \neq  To2.  }
\prule{distinct(Frm1, To1,Frm2, To2)} {Frm1 \neq  Frm2. }
\prule{select\_larger(X, Y , X)}{ \bk1  X  \geq Y.  }
\prule{select\_larger(X , Y, Y) }{  \bk1 Y > X.}
\eldl
\end{example}

As demonstrated by these examples,
XY-stratified programs  allow an
efficient logic-based expression of
procedural algorithms. For instance, the
alternating fixpoint procedure used in the
computation of well-founded models can also
be expressed using these programs \cite{Kemp95}.
In general, XY-stratified programs are quite powerful,
as demonstrated by fact that these programs
(without choice, aggregates, and function symbol) are known to be equivalent to
Statelog programs \cite{LaLuM98} , which have {\sc pspace} complexity and
can express the {\sc while} queries \cite{AHV95}.
Finally, observe that
the bistate programs for the examples used here are nonrecursive.
In general, by making the
computation of the recursive predicate
explicit as it was done for the {\tt anc} example,
it is possible to rewrite  an XY-stratified program $P$ whose
bistate version $P_{bis}$ is recursive into an XY-stratified program
$P'$ whose bistate version $P'_{bis}$ is nonrecursive.

\subsection*{Choice and Aggregates in XY-stratified Programs}

As described in Section \ref{sec:choice}, choice can be
used in stratified programs with no restriction, and
its stable model can be computed by an iterated choice
fixpoint procedure.
Generalizing such notion, the  {\LDL}++ system supports the use
of choice in programs that are XY-stratified
with respect to negation.
The following conditions are
however enforced to assure the
existence of stable models 
for a given program $P$ \cite{GiaMa98}:

\begin{itemize}
\item
The program obtained from $P$ by removing its
choice goals is XY-stratified  w.r.t. negation, and
\inv
\item
If $r$ is a recursive choice rule in $P$, then {\em some} choice
goal of $r$ contains  $r$'s temporal variable
in its left side.
\end{itemize}

After checking these conditions, the {\LDL}++ compiler constructs
$syncbi(P)$ by dropping the temporal
variable from the choice goals and transforming the rest
of the rules as described in the previous section. Then, the
program $syncbi(P)$ so obtained is a stratified choice program and
its stable models can be computed accordingly;
therefore, each stable model for the original XY-stratified program $P$
is computed by simply applying
Procedure~\ref{xycomp} with no modification \cite{ZaCe97,GiaMa98}.

Using the simple syntactic characterization given in Section~\ref{sec:udas},
{\LDL}++ draws a sharp distinction between monotonic and nonmonotonic
aggregates. No restriction is imposed
on programs with only monotonic
aggregates and no negation. But recursive programs with nonmonotonic
aggregates must satisfy the following
conditions (which assure that
once the aggregates are expanded as described in Section~\ref{sec:udas}
the resulting choice program satisfies the XY-stratification
conditions for choice programs discussed in the previous paragraph):

\inv
\begin{itemize}
\inv \item
For each recursive rule, the temporal variable
must be contained in the group-by attributes.
\item
The bistate version of $P$ must be stratified w.r.t. negation
and nonmonotonic aggregates, and
\end{itemize}
After checking these simple conditions, the {\LDL}++ compiler
proceeds with the usual computation of  $syncbi(P)$ as
previously described.

For instance, the following XY-stratified program with aggregates
expresses Floyd's algorithm to compute the least-cost path between pairs of
nodes in a graph. Here,
$\tt g(X,Y,C)$ denotes an arc from $\tt X$ to
$\tt Y$ of cost $\tt C$:
\begin{example}{Floyd's least-cost paths between all node pairs.}

\bldl
\prule{delta(0, X, Y, C) }{    g(X,Y,C).}
\prule{new(J+1, X, Z, C) }{ delta(J,X,Y,C1), all(J,Y,Z,C2), C=C1+C2.}
\prule{new(J+1, X, Z, C) }{ all(J,X,Y,C1), delta(J,Y,Z,C2), C=C1+C2.}
\pfact{newmin(J, X, Z, min\<C \>) \leftarrow \hspace*{-1.2cm}}
{\hspace*{1.2cm} new(J, X, Z, C).}
\prule{discard(J, X, Z, C) }{newmin(J, X, Z, C1), all(J, X, Z, C2),
C1 \geq C2.}
\prule{delta(J, X, Z, C) }{newmin(J, X, X, C), \neg discard(J, X, Z, ).}
\prule{all(J+1, X, Z, C) }{all(J,X,Z,C), \neg delta(J+1, X, Z, \_).}
\prule{all(J, X, Z, C)  }{delta(J, X, Z, C).}
\eldl
\end{example}
The fourth rule in this
example uses a nonmonotonic min aggregate to select the least
cost pairs among those just generated (observe that the temporal
variable $\tt J$ appears among the group-by attributes).
The next two rules derive the new $\tt delta$ pairs by discarding from
$\tt new$ those that are larger than any existing pair in $\tt all$.
This new $\tt delta$ is then used to update $\tt all$ and compute
new pairs.

By supporting UDAs, choice, and XY-stratification
{\LDL}++ provides a powerful, fully integrated framework
for expressing logic-based computation and modelling.
In addition to express complex computations~\cite{ldldemo},
this power has been used to model the AI planning problem \cite{BSZ97},
database updates, and active database rules~\cite{Zan97}.
For instance, to model AI planning,
preconditions can simply be
expressed by rules, choice can be used to select among applicable
actions, and frame axioms can be expressed by
XY-stratified rules that describe changes from the old state to
the new state \cite{BSZ97}.

\section{The System}

The main objectives in the design of the {\LDL}++ system,
were (i) strengthening the  architecture of the previous {\LDL}
system~\cite{Ceta}, (ii)
improving the system's usability and the application
development turnaround time, and
(iii) provide efficient support for the new language constructs.

While the
first objective could be achieved by building on
and extending the general
architecture of the predecessor {\LDL} system, the second objective
forced us to depart significantly from the compilation
and execution approach used by the {\LDL} system.  In fact, the old system
adhered closely to the set-oriented semantics of
relational algebra and relational databases; therefore, it
computed and accumulated all partial results before returning the
whole set to the user. However, our experience in developing applications
indicated that a more interactive and incremental
computation model was
preferable: i.e., one where users
see the results incrementally as they are produced. This
allows developers to  monitor  better the computation as it
progresses, helping them debugging their programs,
and, e.g., allowing them to stop promptly
executions  that have fallen into
infinite loops.

Therefore, {\LDL}++ uses a pipelined
execution model, whereby tuples are generated one
at a time as they are needed (i.e., lazily as
the consumer requests them, rather than eagerly).
This approach also realizes objective (iii) by providing better support
for new constructs, such as
choice and on-line aggregation,
and for intelligent backtracking optimization
(discussed in the next section).

The {\LDL}++ system also adopted a shallow-compilation approach
that achieves faster turnaround during program development and enhances
the overall usability; this approach also made it easier
to support on-line debugging and
meta-level extensions. The previous {\LDL} system
was instead optimized for performance; thus, it used a
deep-compilation approach where the original program was translated into
a (large) C program---whose compilation and linking slowed the
development turnaround time. The architecture of the system is
summarized in the next section; additional information, a web
demo, and instructions on downloading for noncommercial use
can be found in~\cite{ldldemo}.
\subsection{Architecture}

The overall architecture of the {\LDL++} system and its main
components are shown in Figure~1.  The
major components of the system are:

\paragraph{The Compiler}

The compiler reads in {\LDL}++ programs and constructs
the {\em Global Predicate Connection Graph} (PCG). For each query form, the
compiler partially evaluates the PCG, transforming it into a network of
objects that are executed by the interpreter. The
compiler is basically similar to that of the old system \cite{Ceta},
and is responsible for checking the safety of queries,
and rewriting the recursive rules
using techniques such the Magic Sets method
\cite{BMSU86}, and the more specialized  methods for
left-linear and right-linear rules \cite{Ullm}.
These rewriting techniques result
in an efficient execution plan
for queries.
\begin{figure}[t]
\includegraphics[height=9.0cm]{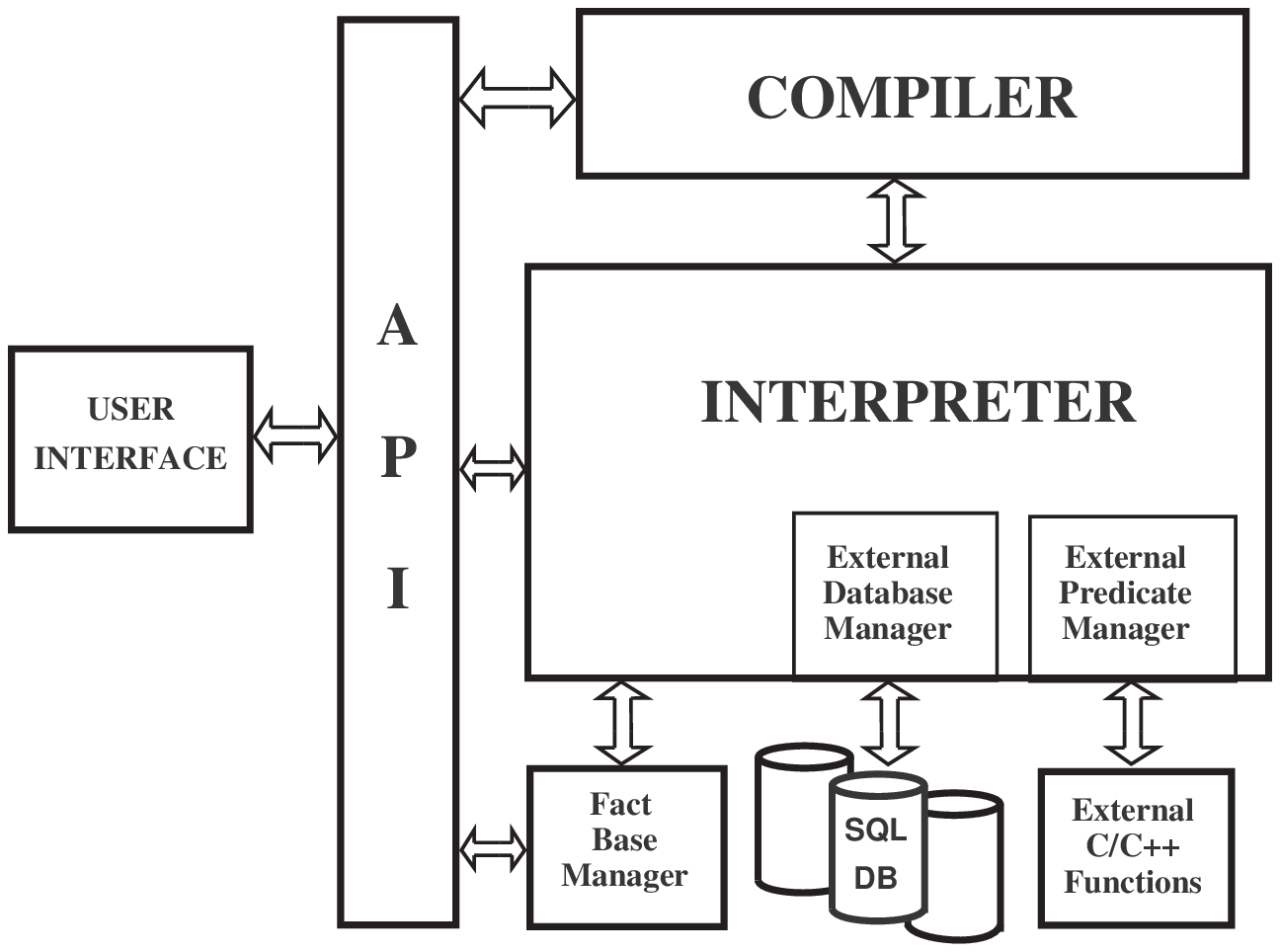}
\caption{{\LDL}++ Open Architecture}
\label{ldl_arch}
\end{figure}






\paragraph{The Database Managers }

The {\LDL} experience confirmed the desirability supporting access to (i)
an internal (fast-path) database and (ii) multiple external DBMSs in a
transparent fashion. This led to the design of a new system where the two
types of database managers are fully integrated.

The internal database is shown in Figure  \ref{ldl_arch}
as Fact Base Manager. This module supports
the management and retrieval of {\LDL}++
complex objects, including sets and lists, and of temporary relations obtained
during the computation.  In addition to supporting users' data
defined by the schema as internal relations, the interpreter relies on
the local
database to store and manage temporary data sets.  The internal database is
designed as a virtual-memory record manager: thus its internal
organization and
indexing schemes
are optimized for the situation where the pages containing frequently
used data can reside in main
memory. Data is written back onto disk at the commit point of each update
transaction; when the transaction aborts the old data is instead
restored from disk.

The system also supports an
external database manager, which is designed to optimize access to
external SQL databases; this is described in Section \ref{sec:externaldb}.

\paragraph{Interpreter}

The interpreter receives as input a graph of executable objects corresponding
to an {\LDL}++ query form generated by the compiler, and executes it by
issuing get-next, and other calls, to the local database. Similar calls are
also issued by the External Database Manager
and the External Predicate Manager  to, respectively,
external databases,  and external functions or software
packages that follow the C/C++ calling conventions. Details on
the interpreter are presented in the next section.

\paragraph{User Interface}

All applications written in C/C++ can call the
{\LDL}++ system via a standard API;  thus
applications written in {\LDL++} can be embedded in other procedural
systems.

One such application  is a line-oriented command interpreter
supporting a set of predefined user commands, command completion
and on-line help. The command interpreter is supplied with the
system, although it is not part of the core system. Basically, the
interface is an application built in C++ that can be replaced with
other front-ends, including graphical ones based on a GUI, without
requiring any changes to the internals of the system. In
particular, a Java-based interface for remote demoing was added
recently \cite{ldldemo}.

\subsection{Execution Model and Interpreter}

The abstract machine for the {\LDL}++ interpreter is based upon the
architecture described in \cite{CGK89}.  An {\LDL}++ program is transformed
into a network of active objects and the graph-based interpreter then
processes these objects.

\paragraph{Code generation and execution}

Given a query form, an {\LDL}++ program is transformed into a Predicate
Connection Graph (PCG), which can be viewed as an AND/OR graph with
annotations. An OR-node represents a predicate occurrence and each AND node
represents the head of a rule.  The PCG is subsequently compiled into an
evaluable data structure called a LAM (for {\LDL}++ Abstract Machine),
whose nodes are implemented as
instances of C++ classes.  Arguments are passed
from one node to the other by means of variables. Unification is
done at compile time and the sharing of variables avoids useless assignments.

Each node of the generated LAM structure has a virtual\footnote{Similar to a
C++ virtual function} ``GetTuple'' interface, which evaluates the
corresponding predicate in the program. Each node also stores a state variable
that determines whether this node is being ``entered'' or is being
``backtracked'' into. The implementation of this ``GetTuple'' interface
depends on the type of node.  The most basic C++ classes are OR-nodes and
AND-nodes; then there are several more specialized subclasses of
these two basic types. Such
subclasses include the special OR-node
that serves as the distinguished {\em root}
node for the query form,
internal relations AND-nodes, external relations
AND-nodes, etc.

\paragraph{And/OR Graph}

For a generic OR node corresponding to a derived relation, the ``GetTuple''
interface merely issues ``GetTuple'' calls to
its children (AND nodes).
Each successful invocation automatically instantiates the
variables of both the child (AND node) and the parent (OR node). Upon
backtracking, the last AND node which was successfully executed is executed
again. The ``GetTuple'' on an OR node fails when its last AND node child
fails.

The {\em Dataflow points} represent different entries into the AND/OR nodes, each
entry corresponding to a different state of the computation. The dataflow
points associated with each node are shown in the
following table (observe their similarity to ports in Byrd's Prolog
execution model \cite{BYRD80}):\\

\begin{tabular}{lll}
\hline\hline
DATAFLOW & POINT & STATE  OF  COMPUTATION \\ \hline
{\sc entry}   & {\em e\_dest}         & getting first tuple of node \\
{\sc backtrack} & {\em b\_dest}      & getting next tuple of node \\
{\sc success}  & {\em s\_dest}       & a tuple has been generated \\
{\sc fail} & {\em f\_dest}  & no more tuples can be generated \\\hline\hline\\
\end{tabular}

A dataflow point of a node can be directed to a dataflow point of a different
node by a {\em dataflow destination}.  The {\em entry destination (e\_dest)}
of a given node is the dataflow point to which its entry point is directed.
Similarly, {\em backtrack (b\_dest), success (s\_dest)}, and {\em fail
destinations (f\_dest)} can be defined. The dataflow destinations represent
logical operations between the nodes involved; for example a join or union of
the two nodes. The dataflow points and destinations of a node describe how the
tuples of that node are combined with tuples from other nodes (but not how
those tuples are generated).

To obtain the first tuple of an OR node we get
the first tuple of its first
child AND node. To obtain the next tuple from an OR node we request
it from the AND node that generated the previous tuple.
Observe that the currently ``active''
AND node must be determined at run-time.
When no more tuples can be
generated for a given AND node, then we go to the next AND
node, till the last child AND node
is reached (At this point no more tuples can be
generated for the OR node). Thus, we have:

\begin{tabbing}
\small
\hspace{1.4cm}{\bf OR nodes:} \= {\em e\_dest}: \= the e\_dest of the first child AND-node \\
               \> {\em b\_dest}: \> the b\_dest of  the ``active'' child AND node\\
               \> {\em f\_dest}: \> {\bf if} \=node is first OR node in rule\\
               \>            \> {\bf then} the f\_dest point of parent AND node \\
               \>            \> {\bf else} the b\_dest of previous OR node\\
               \> {\em s\_dest}: \> {\bf if} node is last OR node in a rule\\
               \>            \> {\bf then} the s\_dest of parent AND node \\
               \>            \> {\bf else} the e\_dest of next OR node.
\end{tabbing}

The execution of an AND node is conceptually less complicated.  Intuitively,
the execution corresponds to a nested loop, where, for each tuple of the first
OR node, we generate all matching tuples from the next OR node. This continues
until we reach the last OR node. Thus, when generating the next tuple of an
AND node, we generate the next matching tuple from the last OR node. If there
are no more matching tuples, we generate the next tuple from the previous OR
node. When there are no more tuples to be generated by the first OR node, we
can generate no more tuples for the AND node.
Thus we have:

\begin{tabbing}
\hspace{1.4cm}{\bf AND nodes:} \= {\em e\_dest}: \= the e\_dest of first OR child \\
                \> {\em b\_dest}: \> the b\_dest of last OR child \\
                \> {\em f\_dest}: \> {\bf if} \= node is last AND child \\
                \>            \> {\bf then} f\_dest of parent OR node \\
                \>            \> {\bf else} e\_ dest of next AND node \\
                \> {\em s\_dest}: \> s\_dest of parent OR node.
\end{tabbing}

Given a query, the {\LDL}++ system first
finds the appropriate LAM graph
for the matching query form, then stores any constant
being passed to the query form by initializing the variables attached to the
{\em root} node of the LAM graph. Finally, the system begins the
execution by repeatedly calling the
``GetTuple'' method on the root of this graph.
When the call fails the execution is complete.

\paragraph{Lazy Evaluation of Fixpoints}

{\LDL}++ adopts a lazy evaluation approach ({\em pipelining}) as its primary
execution model, which is naturally supported by
the  AND/OR graph described above. This model is also supported
through the lazy evaluation of fixpoints. The traditional implementation
of fixpoints \cite{Ullm,ZaCe97} assumes an eager
computation where new tuples are generated till the fixpoint is reached.
 {\LDL}++ instead supports lazy computation
where the recursive rules produce new tuples only in response to
the goal that, as a consumer, calls the recursive predicate.
Multiple consumers can be served by one producer, since each consumer
 $j$ uses a separate cursor $C_j$ to access the relation $R$
written by the producer.  Whenever $j$ needs a new tuple, it proceeds as
shown in Figure \ref{lazy}.

\begin{figure}[b]
\caption{\bf Lazy Fixpoint Producer}
\label{lazy}
\begin{description}
\item
[~~~Step 1.] Move the cursor $C_j$ to the next tuple  of $R$,
and consume the tuple.
\item
[~~~Step 2.] If Step 1 fails (thus, $C_j$ is the last tuple of
$R$), check the fixpoint flag $F$.
\item
[~~~Step 3.] If the fixpoint is reached, return failure.
\item
[~~~Step 4.] If the fixpoint is not reached, call the current rule to generate
a new tuple.
\item
[~~~Step 5.] If a new tuple is generated, add it to the relation $R$, advance
$C_j$ and return the tuple.
\item
[~~~Step 6.] Otherwise, repeat Step 2. \\
\end{description}
\end{figure}

A limitation of pipelining is that the internal state of each node must be
kept for computation to resume where the last call left off.  This creates a
problem when several goals call the same predicate (i.e. the same subtree in the
PCG is shared). Multiple invocations of a shared node can interfere with each
other (non-reentrant code).  Solutions to this problem
include (i) using a stack as in Prolog, and (ii) duplicating
the source code as in the {\LDL} system---thus
ensuring that the PCG is a tree,
rather than a DAG~\cite{Ceta}. In the
{\LDL}++ system, we instead  use the lazy producer
approach  described above
for situations where the calling goals have no bound argument.
If there are bound arguments in consuming predicates we
duplicate the node.  However, since each node is implemented as a C++ class,
we simply generate multiple instances of this class---i.e., we
duplicate the data
but still share the code.

\paragraph{Intelligent Backtracking}
Pipelining makes it easy to implement optimizations such
as existential optimization
and intelligent backtracking
\cite{Ceta}.
Take for instance the following example:

\begin{example} Intelligent Backtracking.
\inv
\bldl
\prule {query3(A, B)} {b1(A), p(A, B), b2(A).}
\eldl
\end{example}

Take the situation where the first $\tt A$-value generated by $\tt b1$ is
passed to $\tt p(A,B)$, which succeeds and passes the value of $\tt A$ to $\tt
b2$. If the first call to this third goal fails, there is no point in going
back to $\tt p$, since this can only return a new value for $\tt B$. Instead,
we have to jump back to $\tt b1$ for a new value of $\tt A$. In an eager
approach, all the $\tt B$-values corresponding to each $\tt A$ are computed,
even when they cannot satisfy $\tt b2$.

Similar optimizations were also supported in {\LDL} \cite{Ceta}, but
with various limitations: e.g., existential optimization was
not applied to recursive predicates, since these were not pipelined.
In  {\LDL}++, the techniques are applied uniformly,
since pipelining is now used in the computation of all predicates,
including recursive ones.

\subsection{External Databases}
\label{sec:externaldb}


A most useful feature of the {\LDLXX} system is that it supports
convenient and efficient access to external databases.  As shown in Figure
\ref{ldl_arch}, the External Database Interface (EDI) provides the capability
to interact with external databases. The system is equipped with a generic SQL
interface as well as an object-oriented design that allows easy access to
external database systems from different vendors. To link the system with a
specific external database, it is only necessary to write a small amount of
code to implement vendor-specific drivers to handle data conversion and local
SQL dialects.
The current {\LDLXX} system can link directly with Sybase, Oracle, DB2, and
indirectly with other databases via JDBC
\footnote{Sybase is a trademark of Sybase Inc., Oracle
is a trademark of Oracle Inc., DB2 is a trademark of IBM Inc.}.

The rules in a program make no distinction between internal and external
relations. Relations from external SQL databases are declared in the {\LDLXX}
schema just like internal relations, with the additional specification of the
type and the name of the SQL server holding the data.  As a result, these
external resources are transparent to the inference engine, and applications
can access different databases without changes.  The EDI can also access data
stored in files.

The following example shows the {\LDLXX} schema declarations
used to access an
external relation {\tt employee} in the database {\tt payroll} running
on the server {\tt sybase\_tarski}.

\begin{example} Schema Declaration to external Sybase server.
\begin{verbatim}
database({
        sybase::employee(NAME:char(30),SALARY:int, MANAGER:char(30))
                from sybase_tarski
                use payroll
                user_name 'john'
                application_name 'downsizing'
                interface_filename '/tmp/ldl++/demo/interfaces'
                password nhoj
          } ).
\end{verbatim}
\end{example}

\noindent
The {\LDLXX} system generates SQL queries
that off-loads to the external database server
the computation of (i) the
join, select, project  queries corresponding to
positive rule goals, (ii) the
set differences corresponding to the negated
goals,  and (iii) the aggregate
operations specified in the heads of the rules.

In the following example the rule defines expensive employees as those who
make over 75,000 and more than their managers:

\begin{example} SQL Generation
\begin{verbatim}
   expensive_employee(Name) <-
                employee(Name, Salary1, Manager),
                Salary1 > 75000,
                employee(Manager, Salary2, _),
                Salary1 > Salary2.
\end{verbatim}
\end{example}

\noindent
The {\LDLXX} compiler collapses all the
goals of this rule  and transforms
it into the following SQL node:

\begin{verbatim}
        expensive_employee(Name) <- sql_node(Name).
\end{verbatim}

\noindent
where $\tt sql\_node$ denotes the
following SQL query sent to external database
server:

\begin{verbatim}
        SELECT  employee_0.NAME
        FROM    employee employee_0, employee employee_1
        WHERE   employee_0.SALARY > 75000 AND
              employee_1.NAME = employee_0.MANAGER AND
              employee_0.SALARY > employee_1.SALARY
\end{verbatim}

\noindent
Consequently, access to the external database via {\LDLXX} is as efficient as
for queries written directly in SQL. Rules with negated goals are also
supported and implemented via the {\tt NOT EXIST} construct of SQL.  The
{\LDLXX} \ SQL interface also supports updates to external databases,
including set-oriented updates with qualification conditions.  Updates to
external relations follow the same syntax and semantics as the updates to
local relations. The execution of each query form is viewed as a new
transaction: either it reaches its commit point or the transaction is aborted.

To better support middleware applications,
the coupling of {\LDL}++ with external databases
was further enhanced as follows:

\begin{itemize}
\item {\em Literal Collapsing}: The goals in the body of a rule are
reordered to
ensure that several goals using database relations
can now be supported as a single SQL subquery to be
offloaded to the DBMS.
\item {\em Rule compression}: To offload more complex and powerful queries
the remote database, literals from multiple levels of rules
are combined and the rules are  compressed vertically.
\item {\em Aggregates}:
Rules that contain standard SQL aggregates in their heads can also
be offloaded to the remote SQL system.
\end{itemize}
%
\subsection{Procedural Language Interface}
\label{sec:udps}

As shown in Figure
\ref{ldl_arch},
the {\LDLXX} system is designed to achieve
an open architecture where
links with procedural languages, such C/C++,
can be established in two ways:

\begin{itemize}
\item Via the {Application Programming Interface (API)} which
allows applications to drive the system, and
\item Via the {External Predicate Manager} which allows C/C++
functions to be imported into the inference engine as external predicates.
\end{itemize}

Via the API, any C/C++ routine can call the {\LDLXX} inference engine. The API
provides a set of functions that enable applications to
instruct the {\LDLXX} engine to load a schema, load rules, compile
query forms, send queries, and retrieve results.


Via the external predicate manager,
function defined in C/C++ can be imported into {\LDL}++ and
treated as logical predicates callable as rule goals.
A library of C/C++ functions is also provided to facilitate
the manipulation of internal {\LDL}++ objects,
and the return of multiple answers
by the external functions. Therefore, external functions can have the
same behavior as internal predicates in all aspects, including
flow of control  and backtracking.
Details on these interfaces
can be found in \cite{ldldemo}.

\section{Applications}

The deployment of the {\LDL} and {\LDL}++ prototypes in various real-life
applications have much contributed to understanding the advantages and
limitations of deductive databases in key application domains \cite{Ts1,Ts2}.
Moreover, this experience with application problems,
has greatly influenced the design of the
{\LDL++} system and its successive improvements.

\paragraph{Recursive Queries.}

Our first focus was to compute transitive closures and to
solve various graph problems requiring recursive queries, such as
Bill-of-Materials~\cite{ZaCe97}.
Unfortunately, many of
these applications also require that
set-aggregates, such as counts and minima,
be computed during  the
recursive traversal of the graph. Therefore, these applications could not be
expressed in {\LDL} which only supported stratified semantics, and
thus disallowed the use of negation and aggregation within recursive
cliques. Going beyond stratification thus became a major design
objective for {\LDL}++.

\paragraph{Rapid Prototyping of Information Systems.}

Rapid prototyping from E-R specifications has frequently been suggested as the
solution for the productivity bottleneck in information system design. 
Deductive databases provide a rule-based language for encoding
executable specifications, that is preferable to Prolog and
4GL systems used in the past, because their completely declarative semantics
provides a better basis for specifications and formal methods.
Indeed, {\LDL} proved to be the tool of choice
in the rapid {\em Prototyping of Information Systems} in conjunction with a
structured-design methodology called
{\em POS} (Process, Object and State) \cite{Aeta,Tryon}.
Our proof-of-concept
experiment confirmed the great potential of
deductive databases for the rapid prototyping
of information systems; but this also showed the  need for
a richer environment that also supports
prototyping of graphical interfaces,
and the use of  E-R based CASE tools.  A large
investment in producing such tools
is probably needed before this application
area can produce a commercial success for deductive databases.
\paragraph{Middleware}
At MCC, {\LDL}++ was used in the CARNOT/INFOSLEUTH project to
support semantic agents that carry out distributed, coordinated
queries over a network of databases~\cite{Ong95}. In particular,
{\LDL}++ was used to implement the ontology-driven mapping between
different schemas; the main functions performed by {\LDL}++
include (i) transforming conceptual requests by users  into a
collection of cooperating queries, (ii) performing the needed data
conversion, and (iii) offloading to SQL statements executable on
local schemas (for both relational and O-O databases).

\paragraph{Scientific Databases}

The {\LDL}++ system provided a sound environment on which to experiment
with next-generation database applications, e.g., to support
domain science research, where complex
data objects and novel query and inferencing capabilities are required.

A first area of interest was molecular biology, where several pilot
applications relating to the Human Genome initiative \cite{Eric} were
developed \cite{OvPrTs,TsOlNa}.  {\LDL}++ rules were also used to
model and support taxonomies and concepts from the biological domain, and
to bridge the gap between high-level scientific models
and low-level experimental data when searching and retrieving
domain information \cite{Ts2}.

A second research area involves geophysical databases for atmospheric and
climatic studies \cite{MSZ}. For instance, there is a need for
detecting and tracking over time and space the evolution of synoptic weather
patterns, such as cyclones. The use of {\LDL}++ afforded the rapid development of
queries requiring sophisticated spatio-temporal reasoning on the geographical
database.
This first prototype was then modified to cope with
the large volume of data required, by off-loading much of the search work to
the underlying database.  Special constructs and operators were
also added to express cyclone queries~\cite{MSZ}.

\paragraph{Knowledge Discovery and Decision Support Applications}
The potential of the {\LDL}++
technology in this important application
area was clear from the start \cite{NaTs}, when our
efforts concentrated
on providing the analyst with powerful tools for the
verification and refinement of scientific hypotheses \cite{Ts1}.
In our early experiments, the expert
would write complex verification rules that were then applied
to the data.
{\LDL}++ proved well-suited
for the  rapid prototyping of these rules, yielding what
became known as the `data dredging' paradigm  \cite{Ts1}.

A more flexible methodology was later developed
combining the deductive rules with inductive tools, such
as classifiers or Bayesian estimation techniques.
A prototype of a system combining both the deductive and inductive methods is
the ``Knowledge Miner" \cite{Shen94},
which was used in the discovery of rules
from a database of chemical process data; {\LDL}++ meta predicates proved very
useful in this experiment~\cite{Shen96}.

Other experiments demonstrated the effectiveness of
the system in performing important auxiliary tasks,
such as data cleaning~\cite{Tsetal,bellcore}.
In these applications, the declarative power of {\LDL}++ is used to
specify the rules that define correct data. These allow
record-by-record verification of data for correctness
but also the identification of {\em sets} of records, whose combination
violates the integrity of the data. Finally, the rules
are used to clean (i.e., correct) inconsistent data.
This capability can either be used prior to the
loading of data into the database,
or during the updating of the data after
loading. This early investigations paved the way for
a major research project discussed next focusing on using {\LDL}++
in datamining applications .

\paragraph{Developing Data Mining Applications}
The results of extensive experiences
with an {\LDL}++ based environment for
knowledge discovery  were reported in~\cite{giannotti99,kdd99}.
The first study~\cite{giannotti99}
describes the experience with a fraud detection
application, while the second one reports on a marketing
application using market basket analysis techniques~\cite{kdd99}.
In both studies, {\LDL}++ proved effective at supporting
the many diverse steps involved in the
KDD process.
In ~\cite{kdd99}, the authors explain the rationale
for their approach and the reasons for their success,
by observing that the process of making decisions requires
the integration of two kinds of activities:
(i) knowledge acquisition from data (inductive reasoning),
and (ii) deductive reasoning about the knowledge thus induced,
using expert rules that characterize the specific business domain.
Activity (i) relies mostly on  datamining functions and algorithms
that extract implicit knowledge from raw data
by performing aggregation and statistical analysis on the database.
A database-oriented rule-based system, such as {\LDL}++,
is effective at driving and integrating the different
tasks involved in (i) and  very effective in activity (ii)
where the results of task (i) are refined, interpreted and integrated with
domain knowledge and business rules characterizing
the specific application.

For instance, association rules derived from market basket analysis are
often too low-level to be directly used for marketing decisions.
Indeed, market analysts seek answers to higher-level questions, such as
``Is the supermarket assortment adequate for the company's target
customer class?" or ``Is a promotional campaign effective in establishing
a desired purchasing habit in the target class of customers?".
{\LDL}++ deductive rules
were used in \cite{kdd99} to drive and  control the overall discovery
process and to refine the raw association rules
produced by datamining algorithms into
knowledge of  interest to the business. For instance,
{\LDL}++ would be used to express queries
such as ``Which rules
survive/decay as one moves up or down the product hierarchy?"
or ``What rules have been effected by the recent promotions" \cite{kdd99}.

The most useful properties of {\LDL}++ mentioned in
these studies ~\cite{giannotti99,kdd99,manco2001}
were flexibility, capability
to adapt to the analyst's needs,
and modularity, i.e., the ability to clearly separate the different
functional  components,
and provide simple interfaces for their integration.
In particular, the user defined aggregates described in Section 2.2
played a pivotal roles in these datamining applications
since datamining functions (performing
the inductive tasks) were modelled
as user-defined aggregates which could then be
conveniently invoked by
the {\LDL}++ rules performing the
deductive tasks~\cite{manco2001}. The performance and scalability
challenge was then addressed by encoding these user-defined
aggregates by means of {\LDL}++
procedural extensions, and, for database
resident data, offloading critical tasks
to the database system containing the data~\cite{manco2001}.


\paragraph{Lessons Learned}
The original motivations for the development of the original {\LDL}
system was the desire to extend relational query languages to
support the development of complete applications, thus eliminating the
impedance mismatch from which  applications using
embedded SQL are now suffering.
In particular,
data intensive expert systems  were
the intended `killer' applications for {\LDL}.
It was believed that such applications
call for combining databases and logic programming into
a rule-based language capable of expressing reasoning, knowledge
representation, and database queries. While the original
application area  failed to generate much commercial demand,
other very promising areas emerged since then. Indeed the
success of {\LDL}++ in several areas is remarkable,
considering that {\LDL}++ is suffering from the
combined drawbacks of (i) being a research prototype (rather than
a supported product), and yet (ii) being subject to severe licensing
limitations. Unless the situation changes and
these two handicaps are removed, the only opportunities
for commercial deployments will come from influencing other systems;
i.e., from system that borrow the {\LDL}++
technology to gain an edge in advanced
application areas, such as datamining and
decision support systems.

\section{Conclusion}
Among the many remarkable projects and prototypes \cite{{RaUl95}}
developed in the field of logic and databases
\cite{Minker96}, the {\LDL}/{\LDL}++ project
occupies a prominent position because the level
and duration of its research endeavor, which
brought together theory, systems, and applications.
By all objective measures, the {\LDL}++ project succeeded in
its research objectives. In particular, the
nondeterministic and nonmonotonic
constructs now supported in {\LDL}++
take declarative logic-based semantics
well beyond stratification in terms of power and  expressivity
(and stratified negation
is already more powerful than SLD-NF).
The {\LDL}++ system supports well the
language and its applications. In particular,
the pipelined execution model
dovetails with constructs such as choice and aggregates (and
incremental answer generation), while the system's open architecture
supports tight coupling with external databases,
JDBC, and other procedural languages.
The merits of the {\LDL}++ technology, and
therefore of deductive databases in the large, have been demonstrated
in several pilot applications---particularly datamining applications.

Although there is no current plan to develop {\LDL}++ commercially,
there remain several exciting opportunities to transfer its logic-oriented
technology to related fields. For instance, the new query and
data manipulation languages for web documents, particularly XML
documents, bear affinity to logic-based rule languages.
Another is the extension to SQL databases of the new constructs
and non-stratified semantics developed for {\LDL}++: in fact,
the use of monotonic aggregates in SQL has already been
explored in \cite{icde2k}.

\subsection*{Acknowledgements}
The authors are grateful to the referees for many suggested improvements.
This work was partially supported by NSF Grant IIS-0070135.


\appendix
\subsection*{Appendix I: Aggregates in Logic}

The expressive power of choice can be used to
provide a formal definition of aggregates in logic.
Say for instance that we want to define the aggregate
{\tt avg} that returns the average of all
$\tt Y$-values that satisfy  $\tt d(Y)$. By
the notation used in {\LDL}~\cite{Ceta},
$CORAL$ \cite{Rama93}, and {\LDL}++,
this computation can be specified by the following rule:

\bldl
\prule{p(avg \< Y \>)} {d(Y).}
\eldl

A logic-based equivalent for this rule is \bldl \prule{p(Y)} {
results(avg, Y).} \eldl where $\tt results(avg, Y)$ is derived from
$\tt d(Y)$ by (i) the chain rules, (ii) the $\tt cagr$ rules and
(iii) the $\tt return$ rules.

The chain rules are those of Example 3
that place the elements of $\tt d(Y)$ into an order-inducing chain.

\bldl
\pfact{ chain(nil, nil). } { }
 \prule{ chain(X,Y) } {chain(\_,X), \ d(Y),}
 \pbody{} {choice((X),(Y)), \ choice((Y),(X)). \ \ \ \ }
\eldl
Now, we can define the $\tt cagr$ rules
to perform the inductive computation by
calling the $\tt single$ and $\tt multi$ rules as follows:
\inv \bldl
\prule {cagr(AgName, Y, New)} {chain(nil, Y), Y \neq nil, single(avg,
Y, New).}
\prule{cagr(AgName, Y2, New)} {chain(Y1, Y2), cagr(AgName,
Y1,  Old), }
\pbody{} {multi(AgName, Y2, Old, New).}
\eldl

Thus, the
$\tt cagr$ rules are used to memorize the previous results, and to
apply (i) {\tt single} to the first element of $\tt d(Y)$ (i.e., for
the pattern $\tt chain(nil, Y)$) and (ii) $\tt multi$ to the
successive elements. The return rules are as follows:

\inv
\bldl
\prule{results(AgName, Yield)} {chain(Y1, Y2), cagr(AgName,  Y1,  Old), }
\pbody{} {ereturn(AgName, Y2, Old, Yield).}
\prule{results(AgName, Yield)}{
chain(X, Y), \neg chain(Y, \_),}
\pbody{} {cagr(AgName, Y, Old),}
\pbody{} {freturn(AgName, Y, Old, Yield).}
\eldl

Therefore, we first compute $\tt chain$, and then
$\tt cagr$ that applies the $\tt single$ and $\tt multi$ rules
to every element in the chain. Concurrently, the first
{\tt results} rule produces all the results that
can be generated by the application of the $\tt ereturn$  rules
to the element in the chain.
The final returns are instead computed by the second
$\tt results$ rule that calls on the $\tt freturn$ rules once
the last element in the chain (i.e., the element without successors)
is detected. The second $\tt results$ rule is the only rule using negation;
in the absence of $\tt freturn$ this rule can be removed
yielding a positive choice program that is monotonic
by  Theorem 2. Thus, every aggregate with only early returns is
{\em monotonic with respect to set containment} and can be used
freely in recursive rules. 

\label{lastpage}
\end{document}